\documentclass[journal]{IEEEtran}
\usepackage{graphicx}
\usepackage{braket}

\ifCLASSINFOpdf
\else
\fi

\begin{document}

\title{Quantum-enhanced metrology without entanglement \\ based on optical cavities with feedback}

\author{Lewis A. Clark, Adam Stokes, M. Mubashir Khan, Gangcheng Wang and Almut Beige
\thanks{The School of Physics and Astronomy, University of Leeds, Leeds LS2 9JT, United Kingdom.}%
\thanks{Manuscript received July 1, 2016.}}

\markboth{Journal of \LaTeX\ Class Files,~Vol.~*, No.~*, Month~2016}%
{Shell \MakeLowercase{\textit{et al.}}: Cavity-based quantum-enhanced metrology}

\maketitle
\IEEEpeerreviewmaketitle

\begin{abstract}
There are a number of different strategies to measure the phase shift between two pathways of light more efficiently than suggested by the standard quantum limit. One way is to use highly entangled photons. Another way is to expose photons to a non-linear or interacting Hamiltonian. This paper emphasises that the conditional dynamics of open quantum systems provides an interesting additional tool for quantum-enhanced metrology. As a concrete example, we review a recent scheme which exploits the conditional dynamics of a laser-driven optical cavity with spontaneous photon emission inside a quantum feedback loop. Deducing information from second-order photon correlation measurements requires neither optical non-linearities nor entangled photons and should therefore be of immediate practical interest.
\end{abstract}


\IEEEpeerreviewmaketitle
 \thispagestyle{empty}
 
\section{Introduction} \label{sec1}

This paper compares different strategies to decrease the uncertainty $\Delta \varphi$ for measurements of an unknown phase shift $\varphi$ between two pathways of light when given a certain amount of resources $N$. Using $N$ independent resources, i.e.~deducing $\varphi$ from a measurement signal $M(\varphi)$ obtained from the dynamics of a generator which is linear in $N$, the scaling of the lower bound of the uncertainty $\Delta \varphi$ of the phase measurement with respect to $N$ is given by the standard quantum limit \cite{review},
\begin{eqnarray} \label{Limitsclass}
\Delta \varphi_{\rm class} &\propto & N^{-0.5} \, .
\end{eqnarray}
However, using for example $N$ highly-entangled photons as a resource, the measurement uncertainty $\Delta \varphi$ can be as low as the Heisenberg limit,  
\begin{eqnarray} \label{Limitsquant}
\Delta \varphi_{\rm quant} &\propto & N^{-1} \, .
\end{eqnarray}
An alternative way of enhancing metrology beyond the standard quantum limit is to expose $N$ incoming single photons to a $\varphi$-dependent Hamiltonian which is no longer linear in $N$ \cite{non-lin}. In this case, the uncertainty $\Delta \varphi$ of the phase measurement scales as
\begin{eqnarray} \label{Limitsnonlin}
\Delta \varphi_{\rm non-lin} & \propto & N^{- 0.5 \, k} \, ,
\end{eqnarray}
where $k$ denotes the order of non-linearity of the interaction Hamiltonian with respect to $N$ or describes the interaction between $k$ phase dependent systems. However, multi-photon entanglement and highly-efficient optical non-linearities are hard to implement experimentally and have not yet become readily available for a wide range of applications.  

In some situations, the length of the measurement process, $t$, and not the number of incoming photons, $N$, is the main resource which we want to constrain \cite{PRA}. If we can write the time $t$ in such cases as $t = N \, \Delta t$ with $\Delta t$ being the length of a short time interval, then the standard quantum limit still coincides with Eq.~(\ref{Limitsclass}). In the following we emphasise that the environmental interactions of open quantum systems naturally result in conditional dynamics, like quantum jumps \cite{Blatt}, which can be exploited for quantum computational tasks \cite{PRL,PRL2,PRL3,Clark}. It is shown that the generators of the conditional dynamics can introduce a non-linear resource-dependence with respect to time \cite{prep}. Applying this observation to quantum metrology provides an additional tool which allows us to break the standard quantum limit (\ref{Limitsclass}) and helps us to design scheme which can be implemented relatively easily.  

As an example, we review a recent quantum-enhanced metrology scheme by Clark {\em et al.}~\cite{PRA} which requires only a laser-driven optical cavity inside a quantum-feedback loop. The proposed setup is feasible with current technology  \cite{Kuhn,Kimble}. Differing from Ref.~\cite{PRA}, this paper does not pay as much attention to the concrete analysis of the proposed scheme. Instead, it focusses its attention on what we believe to be the main mechanisms underlying the observed enhancement. Our findings complement the work of other authors \cite{Braun,Openmet,Kok3}, who also observe that quantum-enhancements may be obtained from quantum correlations and sequential measurements in open quantum systems. 

There are five sections in this paper. Section \ref{sec2} reviews the main theoretical models that are commonly used to describe quantum optical systems with spontaneous photon emission. In Section \ref{sec3}, we discuss how to use the non-linear conditional dynamics of an open quantum system to measure the phase shift $\varphi$ between two pathways of light, using the work of Ref.~\cite{PRA} as an example. Section \ref{sec4} emphasises that the observed quantum enhancement is not unexpected by showing that subsequent measurements on a single quantum system provide at least as much information as single-shot measurements on entangled states. Finally, we summarise our findings in Section \ref{sec5}.

\newpage
\section{The quantum jump dynamics \\ of open quantum systems} \label{sec2}

In this section, we review the modelling of open quantum systems with spontaneous photon emission  \cite{Reset,Molmer,Carmichael,Adam}, thereby emphasising that the conditional dynamics of single quantum trajectories, which is associated with quantum jumps, is naturally non-linear. 

\subsection{The Schr\"odinger equation of closed quantum systems}

As is well known, the dynamics of a closed quantum system in the Schr\"odinger picture is given by the Schr\"odinger equation,
\begin{eqnarray} \label{SE}
|\dot \psi \rangle &=& - {{\rm i} \over \hbar} \, H \, |\psi \rangle \, ,  
\end{eqnarray}
where $|\psi \rangle$ is a state vector and $H$ is the time-independent Hamiltonian and energy observable. Solving the above equation for a given initial state $|\psi (0) \rangle$ yields the state vector $|\psi(t) \rangle$, 
\begin{eqnarray} \label{VNE}
|\psi(t) \rangle &=& \exp \left( - {{\rm i} \over \hbar} H t \right) \, |\psi (0) \rangle \, .
\end{eqnarray}
This shows that the generator for the dynamics of a closed quantum system is linear in time.

Suppose the Hamiltonian $H$ depends on an unknown parameter $\varphi$ which we want to measures as accurately as possible. Then the above closed system dynamics, i.e.~the time evolution of a measurement signal $M(t,\varphi)$, can be used to deduce $\varphi$. The longer the system is observed and the larger the measurement time $t$, the more information about $\varphi$ becomes available and the smaller the uncertainty $\Delta \varphi$ of this measurement can become. However $\Delta \varphi$ is limited by the standard quantum limit (\ref{Limitsclass}) with respect to the resource $N$ which measures time.  

\subsection{Master equations of open quantum systems}

When interacting with an environment, the expectation values of physical observables averaged over an ensemble of quantum systems can no longer be deduced from a single state vector $|\psi \rangle$. The ensemble now needs to be described by a density matrix $\rho$. For Markovian systems, $\rho $ necessarily obeys a master equation of Lindblad form. For example, the master equation of an open quantum system with spontaneous photon emission and only a single decay channel can be written as \cite{Adam}
\begin{eqnarray} \label{master}
\dot \rho &=& {\cal L} \, \rho 
\end{eqnarray}
with the linear superoperator ${\cal L}$ given by 
\begin{eqnarray} \label{master3}
{\cal L} \, \rho &=& - {{\rm i} \over \hbar} \, \left[ H , \rho \right] + {1 \over 2} \Gamma \left( 2 \, L \rho L^\dagger - \left[ L^\dagger L , \rho \right]_+ \right) \, . ~~
\end{eqnarray}
Here $\Gamma$ denotes the spontaneous decay rate and $L$ is the so-called Lindblad operator. In the case of an optical cavity, $L$ simply coincides with the cavity photon annihilation operator $c$ and $\Gamma$ becomes the spontaneous cavity decay rate $\kappa$. In analogy to Eq.~(\ref{VNE}) we now find that
\begin{eqnarray} \label{VNE2}
\rho (t) &=& \exp \left( {\cal L} t \right) \rho(0) \, ,
\end{eqnarray}
where $\rho(0)$ denotes the initial state of the open quantum system. As in the previous subsection, we see that the generator of the system dynamics is linear in time. This is analogously to the situation of closed quantum systems in the previous subsection. The standard quantum limit (\ref{Limitsclass}) still gives a lower bound for the uncertainty $\Delta \varphi$ for the measurement of an unknown parameter $\varphi$ with respect to time. 

\subsection{Unravelling into quantum trajectories} \label{lastnumber}

Having a closer look at microscopic derivations of Eq.~(\ref{master}) (see e.g.~Refs.~\cite{Reset,Molmer,Carmichael,Adam}) shows that Eq.~(\ref{master}) is the result of averaging over all possible quantum trajectories that the quantum system can experience. If the system is continuously monitored and all its photon emission times are known, then it can still be described by a pure state vector at all times. To point out the relevant unravelling of Eq.~(\ref{master}), we now write the derivative $\dot \rho$ as
\begin{eqnarray} \label{master2}
\dot \rho &=& - {{\rm i} \over \hbar} \, \left[ H_{\rm cond} \rho - \rho H_{\rm cond}^\dagger \right] + \Gamma \, L \rho L^\dagger   
\end{eqnarray}
with $H_{\rm cond}$ being the non-Hermitian conditional Hamiltonian given by 
\begin{eqnarray} \label{Hcond}
H_{\rm cond} &=& H - {{\rm i} \over 2} \hbar \Gamma \, L^\dagger L 
\end{eqnarray}
and with $L$ being the same Lindblad operator as in Eq.~(\ref{master3}). Re-writing our master equation in this compact form allows us to identify two subensembles. The first term in Eq.~(\ref{master2}) is the time derivative of the unnormalised density matrix of the subensemble of quantum systems without a photon emission in a small time interval $(t, t+\Delta t)$, while the second term refers to the subensemble experiencing an emission. The normalisation of both terms indicates their relative sizes.

More concretely, under the condition of no photon emission in $(t,t+\Delta t)$, the state vector $|\psi (t) \rangle$ evolves such that
\begin{eqnarray} \label{K0}
|\psi(t + \Delta t) \rangle &=& K_0 \, |\psi(t) \rangle /  \| \, K_0 \, |\psi(t) \rangle \, \|
\end{eqnarray} 
with the operator $K_0$ being the conditional no-photon time evolution operator
\begin{eqnarray} \label{K0more}
K_0 &=& \exp \left( - {{\rm i} \over \hbar} H_{\rm cond} \Delta t \right) \, .
\end{eqnarray} 
The generator of the no-photon time evolution of our open quantum system is again linear in time. However, in case of a photon detection in $(t,t+\Delta t)$, the state vector of the quantum system changes into 
\begin{eqnarray} \label{K1}
|\psi(t + \Delta t) \rangle &=& K_1 \, |\psi(t) \rangle  /  \| \, K_1 \, |\psi(t) \rangle \, \|
\end{eqnarray} 
with the reset operator $K_1$ given by
\begin{eqnarray} \label{K1more}
K_1 &=& \left(\Gamma \Delta t \right)^{1/2} \, L \, .
\end{eqnarray} 
In the case of a photon emission, a so-called quantum jump occurs accompanied by sudden jumps of expectation values \cite{Blatt}. The generators of the conditional dynamics of open quantum systems, which are associated with spontaneous photon emission, are in general highly non-linear due to the constant need for resetting upon emission.

\section{Temporal quantum correlations for quantum metrology} \label{sec3}

Section \ref{sec1} emphasises that generators of dynamics, which are non-linear in the relevant resource, can be used to measure an unknown parameter $\varphi$ with an accuracy $\Delta \varphi$ beyond the standard quantum limit. Section \ref{sec2} shows that the generators for the dynamics of the single quantum trajectories of open quantum systems are in general non-linear. Combining these two observations, one can design novel quantum-enhanced metrology schemes which require neither entanglement nor a non-linear or interacting Hamiltonian. To illustrate this fact we now review a recent proposal by Clark {\em et al.}~\cite{PRA} based on a single optical cavity inside a quantum feedback loop for which time is the resource which we want to constrain. 

\subsection{A quantum-enhanced metrology scheme}

\begin{figure}[t]
\centering
\includegraphics[width=3.5in]{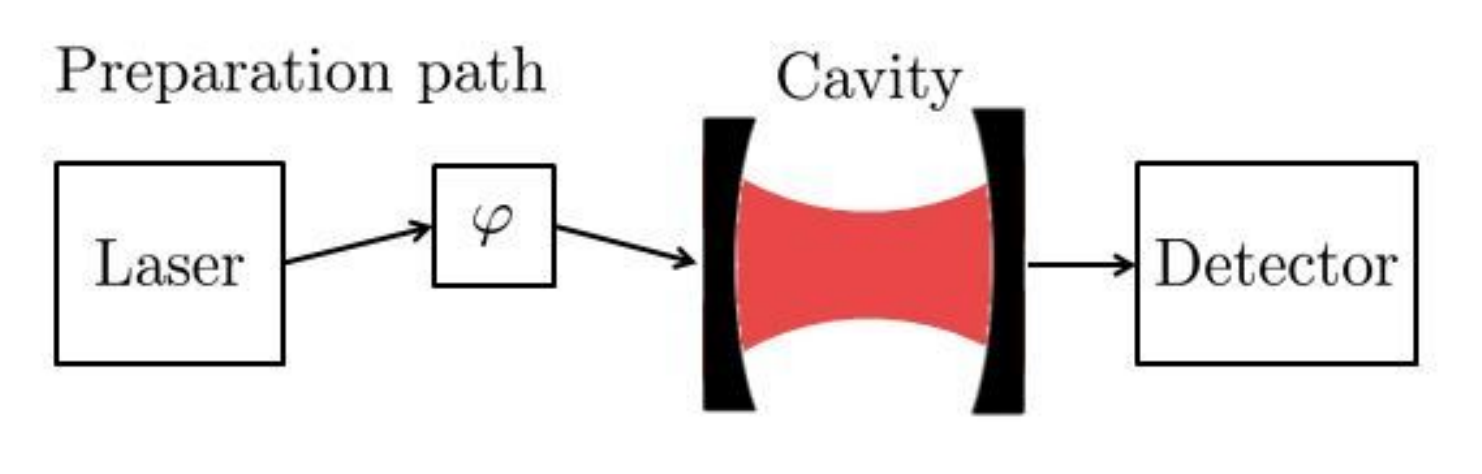}
\caption{Schematic view of the preparation stage. Its purpose is the preparation of the coherent state $|\alpha \rangle$ given in Eq.~(\ref{polar}), which depends on the unknown phase shift $\varphi$. One way of achieving this is to drive a leaky optical cavity with a laser field which experiences $\varphi$ until the system reaches its stationary state.} \label{Stage1}
\end{figure}

\begin{figure}[t]
\centering
\includegraphics[width=3.5in]{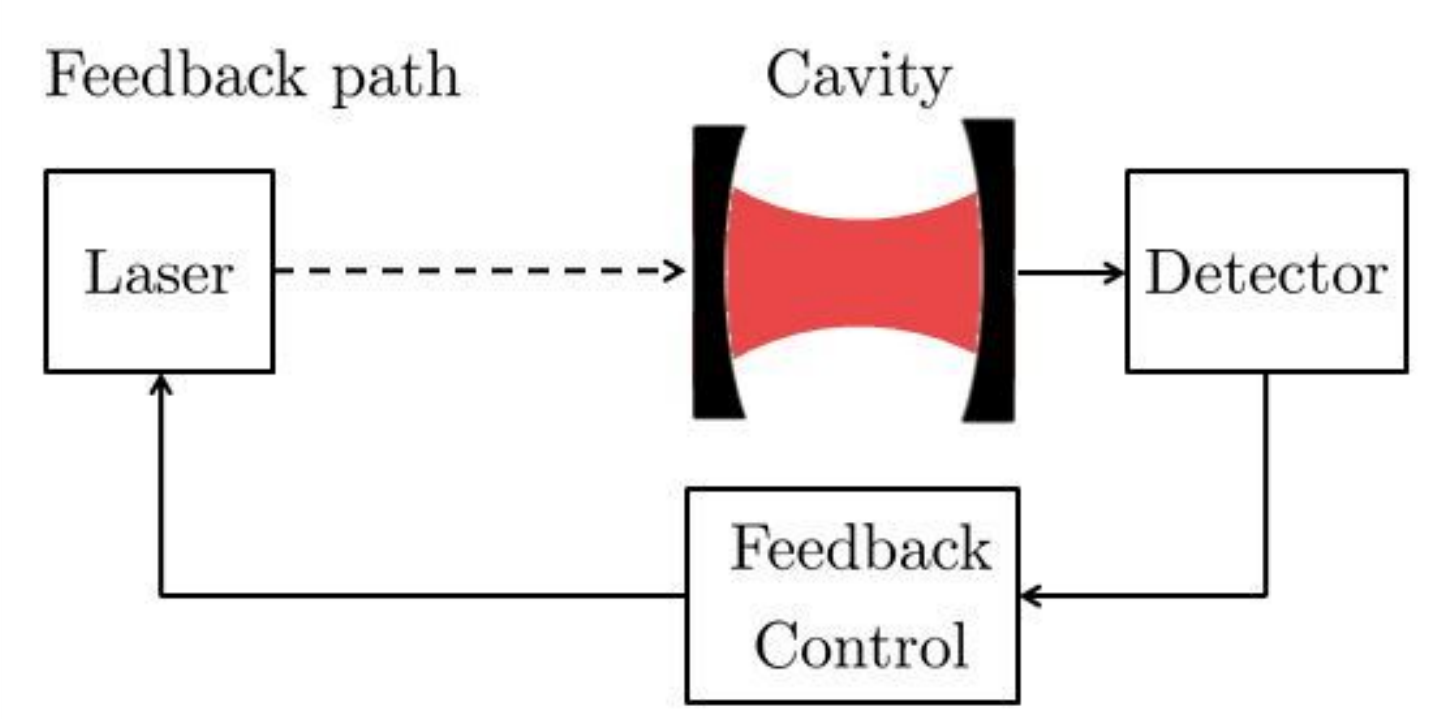}
\caption{Schematic view of the measurement stage. During this stage, the cavity is placed inside a quantum feedback loop. The detection of photon emissions now triggers a laser pulse that displaces the resonator field in a certain direction. The relative phase between the laser field applied during the preparation stage and the quantum feedback laser can be deduced, for example, from second-order photon correlation measurements with an accuracy that increases rapidly in time.} \label{Stage2}
\end{figure}

The quantum metrology scheme in Ref.~\cite{PRA} consists of two main stages, a preparation and a measurement stage. These implement the following tasks:
\begin{enumerate}
\item
Firstly, the preparation stage prepares the cavity field in a coherent state $|\alpha \rangle$ with $\alpha$ being of the form
\begin{eqnarray} \label{polar}
\alpha &=& |\alpha| \, {\rm e}^{{\rm i} \varphi} \, .
\end{eqnarray} 
One way of achieving this is to drive the cavity with a laser field that experiences the phase shift $\varphi$ and to let it relax into its stationary state (c.f.~Fig.~\ref{Stage1}). 
\item Afterwards, during the measurement stage, the cavity is placed inside a quantum feedback loop (c.f.~Fig.~\ref{Stage2}). Whenever a spontaneously emitted photon is detected, a laser pulse is applied. This laser pulse displaces the coherent state inside the resonator in a certain direction, which should be independent of $\varphi$. For simplicity, we assume here that the feedback pulse is approximately instantaneous. 
\end{enumerate}
Since the feedback laser does not experience the unknown phase $\varphi$, it provides a reference frame. What the proposed metrology scheme measures is the relative phase between the laser field applied during the preparation stage and the quantum feedback laser. Alternatively we could choose the preparation laser as the reference frame, since this would not change the dynamics of the system. Although doing so might be less favourable in practical applications, let us assume for the rest of the paper that this is the case for the sake of convenience.

In standard approaches to quantum metrology \cite{review,Q.Met2}, the relevant resource is the number of photons experiencing the unknown phase shift $\varphi$. This is due in part to the nature of interferometric experiments. Usually $N$ photons are passed through an interferometer that contains the unknown phase before being measured at the end. In the following we extract $\varphi$ from the photon statistics of the optical resonator during the measurement stage. Hence in our scheme the main resource is not the number of photons $N$ passing through the setup but the number of observations posed to deduce the photon statistics. This number is directly proportional to the duration of the measurement stage $t$ which we can write as $t = N \, \Delta t$.

\subsection{The non-linear conditional dynamics of an optical cavity with feedback}

As we have seen in Section \ref{sec2}, the expectation values of an open quantum system with spontaneous photon emission averaged over all possible quantum trajectories behave as if they were generated by linear operators. However, to enhance quantum metrology beyond the standard quantum limit without using entanglement, we require our setup to behave as if its dynamics were generated non-linearly. One way of achieving this is to deduce the unknown parameter $\varphi$ from a measurement signal $M(t,\varphi)$ corresponding to a pre-selected subset of quantum trajectories, which involves quantum jumps. In the following we therefore have a closer look at the dynamics of subsets of quantum trajectories of the experimental setup in Fig.~\ref{Stage2}.

For an optical cavity with spontaneous photon emission the Lindblad operator is $L=c$, where $c$ is the usual bosonic photon annihilation operator. This leads to a peculiar effect. If prepared in a coherent state $|\alpha \rangle$, as it is in general the case for a laser-driven optical cavity \cite{PRA}, the spontaneous emission of a photon does not change the field inside the resonator. In other words, there are no quantum jumps in this case. The reason for this is that the coherent states are the eigenstates of the photon annihilation operator $c$. To use the setup in Fig.~\ref{Stage1} nevertheless for quantum-enhanced metrology, we either need to prepare the cavity field in a non-coherent state or we need to replace $L$ by another Lindblad operator.   

A straightforward way of changing $L$ is to introduce quantum feedback. This is why we propose to place the cavity during the measurement stage into a quantum feedback loop, as illustrated in Fig.~\ref{Stage2}. If the feedback operation depends on the unknown parameter $\varphi$, then its introduction moreover results in an effective $\varphi$-dependence of the jump operation. Moreover, changing the state of the cavity upon the detection of photon essentially generates temporal correlations in the conditional dynamics of the quantum system. These correlations mean that the information corresponding to an individual quantum trajectory with respect to two or more different time steps is no longer additive, thereby allowing for scaling beyond the standard quantum limit.

Taking quantum feedback into account in the theoretical model, which we introduced in Section \ref{sec2}, is straightforward. The no-photon time evolution remains the same. However, the operator $K_1$ in Eq.~(\ref{K1more}) now needs to be replaced by $R(\varphi) K_1$ where $R(\varphi)$ is a unitary operator \cite{PRA,Clark,Wiseman-Milburn}. In other words, all we have to do is to replace the Lindblad operator $L$ by another operator $L(\varphi)$,
\begin{eqnarray} \label{pec2}
L &\longrightarrow & L(\varphi)  = R(\varphi) L \, .
\end{eqnarray}
For example, suppose every feedback operation $R(\varphi)$ displaces the coherent state inside the resonator by a certain amount $\beta (\varphi)$ such that
\begin{eqnarray}
R(\varphi) \, |\alpha \rangle &=& |\alpha + \beta (\varphi) \rangle \, .
\end{eqnarray}
Then the detection of a photon results indeed in an effective quantum jump. An effective non-linearity has been created which can be explored for quantum metrology. 

\begin{figure}[t]
\centering
\includegraphics[width=3.6in]{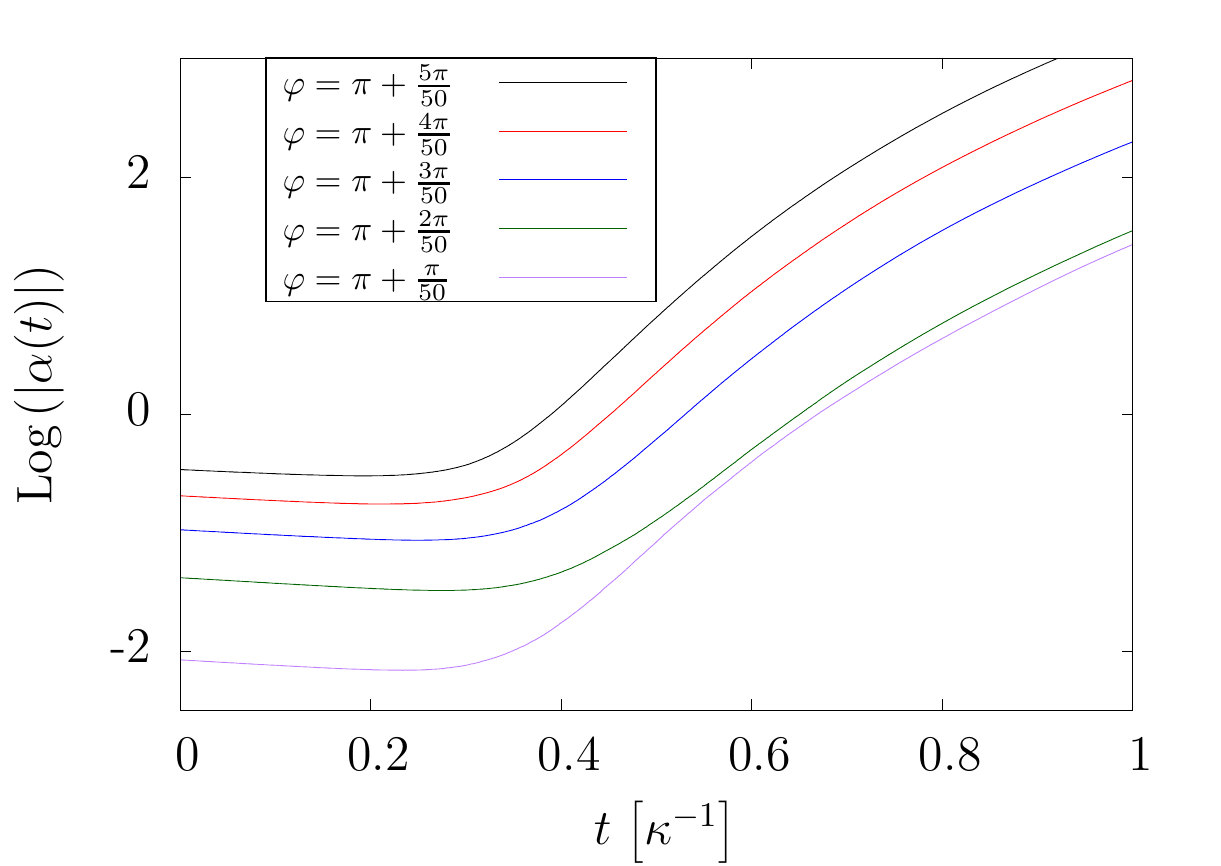}
\caption{
Logarithmic plot of $|\alpha(t)|$ averaged over all the possible quantum trajectories of the subensemsbles with a photon detection and a subsequently applies feedback operation $R(\varphi)$ during the first time step of the measurement stage $(0,\Delta t)$ for five different values $\varphi$ which are all close to $\pi$. The figure is the result of a numerical simulation, which averaged over $10^5$ randomly generated quantum trajectories. Here we assume $\alpha(0)=2$ and $\beta = |\alpha(0)| {\rm e}^{{\rm i} \varphi}$. Moreover, each time step $\Delta t$ is much smaller than the cavity photon life time $1/\kappa$.} \label{Distancedisplacelog}
\end{figure}

Breaking the standard quantum limit in Eq.~(\ref{Limitsclass}) when time is the resource which we want to constrain requires that the dynamics of the cavity is very sensitive to changes of the unknown parameter $\varphi$. In order to be able to distinguish two parameters $\varphi_1$ and $\varphi_2$, the corresponding measurement signals $M(t,\varphi_1)$ and $M(t,\varphi_2)$ need to evolve such that their distance grows non-linearly in time. That this can be the case for the experimental setup which we consider here is illustrated in Fig.~\ref{Distancedisplacelog}. Suppose $\beta = |\alpha(0)| {\rm e}^{{\rm i} \varphi}$ and $\varphi = \pi$. Then the detection of a photon at $t=0$ prepares the cavity in its vacuum state, i.e.~in the coherent state $|\alpha \rangle$ with $\alpha = 0$. Fig.~\ref{Distancedisplacelog} is a logarithmic plot of $|\alpha(t)|$ averaged over all quantum trajectories with a photon detection during the first time step of the measurement stage $(0,\Delta t)$ for different values $\varphi$ which are all close to $\pi$.  All curves separate very quickly from the curve corresponding to $\varphi = \pi$. Since the plot is logarithmic, we see clearly that this happens in a highly non-linear fashion with respect to time.

\subsection{Second order photon correlation functions}

In order to obtain information about $\varphi$ more efficiently than suggested by the standard quantum limit, we need to find a measurement signal $M(t,\varphi)$, which cannot be written as an ensemble average but depends strongly on the appearance of quantum jumps. Taking the discussion in the previous subsection into account, we now have a closer look at the second-order photon correlation function $G^{(2)}(t,t')$, which is given by the joint probability
\begin{eqnarray} \label{G2}
G^{(2)}(t,t') &\equiv & I(t|t') I(t') \, ,
\end{eqnarray}
where $I(t|t')$ denotes the probability for the detection of a photon at a time $t$ conditional on the detection of a photon at $t'$.  Second-order correlation functions are usually normalised by the product of the photon emission rate at $t'$ and at $t$. Doing so and dividing Eq.~(\ref{G2}) by $I(t') I(t)$, we define the renormalised second order photon correlation function, $g^{(2)}(t,t')$, by 
\begin{eqnarray}
g^{(2)}(t,t') &\equiv & {I(t|t') \over I(t)} \, .
\end{eqnarray}
Now $g^{(2)}(t,t')$ depends no longer on the efficiency of the detector shown in Fig.~\ref{Stage2} and can be measured accurately and relatively easily, even when using imperfect single-photon detectors. 

Ref.~\cite{PRA} uses measurements of the second order photon correlation function $g^{(2)}(t,0)$, where $t = N \, \Delta t$ denotes the length of the measurement stage, to deduce information about an unknown phase $\varphi$ between two pathways of light. This means, we propose to measure the joint probability of detecting a photon at the start of the measurement stage and another photon at the end, after a time $t$. For simplicity, we ignore photon emissions between these two points. Nevertheless, we found that the uncertainty $\Delta \varphi$ scales as $N^{-0.71}$, 
\begin{eqnarray} \label{pec}
\Delta \varphi &\propto & N^{-0.71} \, ,
\end{eqnarray}
when $\varphi  = \pi $ and $\beta = |\alpha|$, which surpasses the standard quantum limit \cite{PRA}. Although no entanglement is used, the origin of the quantum enhancement in Eq.~(\ref{pec}) is still of a quantum nature. The second-order correlation function has no classical analogue and its measurement requires the detection of individual photons, although unit detection efficiency is not required.

What might seem most surprising about Eq.~(\ref{pec}) is that the measurement uncertainty $\Delta \varphi$ decreases rapidly, as $t$ increases. The longer one waits, the more information is unvealed about $\varphi$. This again is due to an interesting property of optical cavities inside instantaneous quantum feedback loops. A more detailed analysis of the dynamics of the experimental setup in Fig.~\ref{Stage2} shows that the field inside the resonator never reaches a stationary state \cite{PRA,students} unless when being placed exactly into its vacuum state. Whenever a photon is emitted, a quantum feedback pulse occurs and increases the number of photons inside the resonator. This increases the probability for another photon emission and so on. If the quantum feedback loop is truly instantaneous, the mean number of photons inside the cavity may easily diverge. Hence the longer one waits, the more easily it becomes to distinguish these two scenarios and to determine whether the system is initially in its vacuum state or not. Here this question is equivalent to asking whether $\varphi = \pi$ or not.

Indeed the quantum enhanced-metrology scheme in Ref.~\cite{PRA} exploits the fact that the cavity possesses two different types of dynamics which separate in time. Similar effects have been studied for example in Ref.~\cite{Openmet} for quantum metrology applications. Due to the effective infinite dimensional Hilbert space of the field inside an optical cavity, the divergence between both types of dynamics may become arbitrarily large in principle, thereby allowing for the scaling in Eq.~(\ref{pec}) to be preserved for an indefinite time, which is not the case for the scenarios studied in Ref.~\cite{Openmet}.   

\section{Temporal quantum correlations and entanglement} \label{sec4}

Suppose subsequent generalised measurements are performed on a two-dimensional quantum system prepared in $|\psi \rangle$.  Moreover we assume that these measurements can be described by two Kraus operators $K_0$ and $K_1$ of the form
\begin{eqnarray} \label{Kraus2}
K_i &=& |\tilde \xi_i \rangle \langle \xi_i | \, ,
\end{eqnarray}
where $|\xi_0 \rangle$ and $|\xi_1 \rangle$ are two orthogonal states. However, no such constraint applies to $|\tilde \xi_0 \rangle$ and $|\tilde \xi_1 \rangle$. In case of two measurements, the quantum system changes such that  
\begin{eqnarray}\label{psi}
|\psi \rangle &\rightarrow & \left\{ \begin{array}{l} K_0 \, |\psi \rangle \rightarrow \left\{ \begin{array}{l} K_0 K_0 \, |\psi \rangle \\ K_1 K_0 \, |\psi \rangle \end{array} \right. \\[0.4cm] K_1 \, |\psi \rangle \rightarrow \left\{ \begin{array}{l} K_0 K_1 \, |\psi \rangle \\ K_1 K_1 \, |\psi \rangle \end{array} \right. \end{array} \right. 
\end{eqnarray}
up to normalisation factors. Moreover, suppose we perform a single-shot measurement of $K_0$ and $K_1$ on two quantum systems prepared in an effective state $|\psi_{\rm eff} \rangle$,
\begin{eqnarray} \label{ent}
|\psi_{\rm eff} \rangle &=& \sqrt{p_{00}} \, |\xi_0 |\xi_0 \rangle + \sqrt{p_{01}} \, |\xi_0 \xi_1 \rangle + \sqrt{p_{10}} \, |\xi_1 \xi_0 \rangle  \nonumber \\
&& + \sqrt{p_{11}} \, |\xi_1 \xi_1 \rangle 
\end{eqnarray}
with the coefficients $p_{ij}$ equal to
\begin{eqnarray}
p_{ij} &=& \| K_j K_i \, |\psi \rangle \|^2 \, .
\end{eqnarray} 
One can easily see that both measurements yield the outcome ``$ij$'' with exactly the same probability. The state $|\psi \rangle$ and $|\psi_{\rm eff} \rangle$ have the same information content. However, $|\psi_{\rm eff} \rangle$ is in general an entangled state. For example, if $K_0 = |\xi_1 \rangle \langle \xi_0 |$ and $K_1 = |\xi_0 \rangle \langle \xi_1 |$, we find that 
\begin{eqnarray}
|\psi_{\rm eff} \rangle = \sqrt{p_{01}} \, |\xi_0 \xi_1 \rangle + \sqrt{p_{10}} \, |\xi_1 \xi_0 \rangle \, . 
\end{eqnarray}
Analogously, one can show that $N$ successive measurements on a single system are in general equivalent to a single-shot measurement of $N$ entangled quantum systems.

The quantum-enhanced metrology scheme that we propose in Ref.~\cite{PRA} extracts information about the unknown phase $\varphi$ between two pathways of light by performing $N$ successive measurements on a single quantum system. This means, our scheme is equivalent to performing single-shot measurements on a combination of $N$ entangled quantum systems.  It is therefore not surprising that our scheme can be used to break the standard quantum limit, as the system possesses correlations.  These are correlations between the system and its environment.  It is the measurements upon the environment, i.e. the measurement of photon emission, that accesses these correlations.

\section{Conclusions} \label{sec5}

This paper emphasises that the environmental interactions of open quantum systems with spontaneous photon emission naturally result in non-linear conditional dynamics which can be exploited for quantum metrology and other applications. More concretely, we propose to deduce an unknown parameter $\varphi$ by measuring an expectation value $M(t,\varphi)$ averaged over a subset of preselected quantum trajectories instead of measuring ensemble averages. If the signal $M(t,\varphi)$ evolves with a non-linear generator, the accuracy of the measurement $\Delta \varphi$ can exceed the scaling proposed by the standard quantum limit \cite{prep}. As an example, we reviewed a recent quantum-enhanced metrology scheme which measures the phase shift between two pathways of light using the open system dynamics of the electromagnetic field of an optical cavity inside a quantum feedback loop \cite{PRA}. This scheme should be of immediate practical interest, since it requires neither efficient optical non-linearities nor entangled photons.

\section*{Acknowledgment}

AS and AB acknowledge financial support from the UK EPSRC-funded Oxford Quantum Technology Hub for Networked Quantum Information Technologies NQIT. MMK acknowledges a postdoctoral research fellowship funding from the Higher Education Commission of the Government of Pakistan. GW acknowledges financial support from the NSF of China (Grant No. 11405026) and Government of China through a CSC (Grant No. 201506625070 ).

\ifCLASSOPTIONcaptionsoff
  \newpage
\fi

\end{document}